\documentstyle[12pt,epsf]{article}
\makeatletter

\ifcase \@ptsize
\or % mods for 11 pt
\or % mods for 12 pt
\fi
\advance\textheight by \topskip

\ifcase \@ptsize
\or % mods for 11 pt
\or % mods for 12 pt
\fi

\def\ga{\gamma}         
\def\be{\beta}
\def\al{\alpha}

\def\de{\delta}         
         
\def\sig{\sigma}

\def\an{$a_n^{(1)}$\ }
\def\d4{$d_4^{(1)}$\ }

\def\noi{\noindent}
\def\atft{affine Toda field theory }
\def\begeq{\begin{equation}}
\def\endeq{\end{equation}}
\newcommand{\setR}{\mbox{\sf I\hspace{-0.11em}R}}

\newcommand{\setN}{\mbox{\sf I\hspace{-0.11em}N}}
\newcommand{\setZ}{\mbox{\sf Z\hspace{-0.41em}Z}}
\begin{document}
\renewcommand{\thefootnote}{\fnsymbol{footnote}}

%
%%%%%%%%%%%% reference number as small superscripts %%%%%%%%%%%%%%
%
%
%\def\@cite#1#2{\unskip\nobreak\relax
%    \def\@tempa{$\m@th^{\hbox{\the\scriptfont0 #1}}$}%
%    \futurelet\@tempc\@citexx}
%\def\@citexx{\ifx.\@tempc\let\@tempd=\@citepunct\else
%    \ifx,\@tempc\let\@tempd=\@citepunct\else
%    \let\@tempd=\@tempa\fi\fi\@tempd}
%\def\@citepunct{\@tempc\edef\@sf{\spacefactor=\the\spacefactor\relax}\@tempa
%    \@sf\@gobble}
%
%%%%%%%%%%%%%%%%%%%%%%%%%%%%%%%%%%%%%%%%%%%%%%%%%%%%%%%%%%%%%%%%%%

%\renewcommand{{\baselinestretch}{2}}

\baselineskip 20pt
\def\mb #1	{\mbox{\boldmath$#1$}}
\def\T		{^{^{\!\rm T\!}}}
\def\Insert#1	{\vspace{1in}\begin{center}\framebox{\bf #1 }\end{center}
\vspace{1in}}
\def\tbc	{\Insert{To Be Completed} }
\font\upright=cmu10 scaled\magstep1
\def\stroke{\vrule height8pt width0.4pt depth-0.1pt}
\def\topfleck{\vrule height8pt width0.5pt depth-5.9pt}
\def\botfleck{\vrule height2pt width0.5pt depth0.1pt}
\def\Ctext{{\rlap{\rlap{C}\kern 3.8pt\stroke}\phantom{C}}}
\def\Rtext{\hbox{\upright\rlap{I}\kern 1.7pt R}}
\def\Ntext{\hbox{\upright\rlap{I}\kern 1.7pt N}}
\def\oC{\ifmmode{{\hbox\Ctext}}\else\Ctext\fi}
\def\oN{\ifmmode{\vcenter{\Ntext}}\else\Ntext\fi}
\def\oR{\ifmmode{\vcenter{\Rtext}}\else\Rtext\fi}
\def\epsf#1 {\epsfxsize=\hsize\epsfbox{EPSF/#1 }}
{\pagestyle{empty}
\rightline {DTP-94-35}
\rightline {September 1994}
\rightline {hep-th/9409035}
\vskip 0.9in
\centerline{\LARGE On the Breathers of}
\vskip 0.1in
\centerline{\LARGE \an Affine Toda Field Theory}
\vskip 0.9in
\centerline {\it \Large Uli
Harder\footnote{\tt U.K.F.Harder@durham.ac.uk}, Alexander A.
Iskandar\footnote{\tt A.A.P.Iskandar@durham.ac.uk}\footnote{On leave of
absence from Department of Physics, Institut Teknologi Bandung,
Bandung
40132, Indonesia.}, William A. McGhee}
\centerline {\it Department of Mathematical Sciences,}
\centerline {\it University of Durham,}
\centerline {\it Durham DH1 3LE,}
\centerline {\it England.}
\medbreak
\vskip .5in
\centerline {\Large Abstract}
\medbreak
\noi Explicit constructions of \an \atft breather solutions are
presented. Breathers arise either from two solitons of the same species
or from solitons which are anti-species of each other. In the first case,
breathers carry topological charges. These topological charges lie in the
tensor product representation of the fundamental representations associated
with the topological charges of the constituent solitons. In the second
case, breathers have zero
topological charge. The breather masses are, as expected, less
than the sum of the masses of the constituent solitons.
\vfill
\newpage
}

%
%%%%%%%%%%%%%%%%%%%%%%%%%%% SECTION 1 - INTRODUCTION %%%%%%%%%%%%%%%%%%%%%%%%%%
%

\renewcommand{\theequation}{\thesection.\arabic{equation}}

\setcounter{equation}{0}
\section{Introduction}

Toda field theories have attracted interest for various reasons over the past
few decades \cite{AFZ+MOP,OT3,OT1+2}. Of particular interest is the \atft
which is a perturbed conformal field theory that is still integrable
\cite{HM+EY}. There has been much progress in the understanding of the real
coupling regime of these affine Toda theories
\cite{BCDS1,CM,BCDS2,PED,MF+FLO,BS+SZ,NSL}.

More recently, Hollowood \cite{TH1} showed that the \an series
of affine Toda field theories have complex soliton solutions with real
energy. The reality of energy and momentum has been generalized
by Olive et al. to include all affine Toda theories associated
with the affine Kac-Moody algebras \cite{OTU1,OTU2}.
The sine-Gordon theory is the simplest example of the \an
series, i.e. the $a^{(1)}_{1}$ Toda field theory.
Besides soliton and antisoliton solutions of the sine-Gordon
theory, there also exist oscillating solitonic solutions, the
{\it breathers}. These breathers are bound states of the
sine-Gordon soliton and antisoliton pair.
Since the \an \atft is a generalization of the sine-Gordon theory,
it is natural to ask if such breather solutions also exist for
these theories. Much has been conjectured about the breathers of the
affine Toda field theory \cite{TH1,OTU1,OTU2,FJKO}.
Calculation of the scattering processes of the \an affine Toda
solitons \cite{TH2}
also points to
the existence of these breathers, since there are poles in the soliton
$S$-matrix which correspond to bound states of soliton pairs.
However, an explicit
construction is lacking. The motivation behind this work is to give an
explicit expression of classical breather solutions for the \an \atft
based on a
similar construction to that of the sine-Gordon breather.

The \atft Lagrangian density is given by,
\begeq\label{lagr}
{\cal L} = \frac{1}{2} (\partial_{\mu}\phi)(\partial^{\mu}\phi) -
\frac{m^{2}}{\tilde\be^{2}}\sum_{j=0}^{n} n_{j}(
e^{\tilde\be\al_{j}\cdot\phi} - 1),
\endeq
\noi where $\phi$ is an $n$-dimensional scalar field, $m$ and $\tilde
\be$ are the mass and coupling parameters respectively. The vectors
$\al_{j}$ with $j = 1,2,\ldots,n$ are the simple roots of the Lie
algebra $g$ of rank $n$ associated with the theory. And $\al_{0}$ is
chosen such that the set $\{\al_{0},\al_{j}\}$ represents the
affine Dynkin diagram of an affine Kac-Moody algebras. For the {\it
simply-laced} Kac-Moody algebras, $\al_{0}$ is the negative of the
highest root $\psi = \sum_{j=1}^{n} n_{j}\al_{j}$; in particular, for
the \an series, $n_{j} = 1$ for all $j$. The constants $n_{j}$ are
inserted into
the potential term so that $\phi = 0$ is a vacuum solution.

Substituting the coupling parameter $\tilde\be$ with a
{\it purely imaginary} coupling parameter, $i\be$, the \atft potential
admits multiple vacua. The existence of these vacua signals a
possibility
of topological solitonic solutions which interpolate between them.
The equation of motion can be derived easily from (\ref{lagr}) upon
substitution of the {\it purely imaginary} coupling parameter,
\begeq\label{eom}
\partial^2\phi = -\frac{m^2}{i\be}\sum_{j=0}^{n}
n_{j}\al_{j}\exp(i\be\al_{j}\cdot\phi).
\endeq
\noi Soliton solutions have been calculated using the Hirota's
methods \cite{TH1,MM}, and using an algebraic method
\cite{OTU1,OTU2}. This algebraic
approach is a generalization of the Leznov-Saveliev solution of
the conformal Toda theory \cite{LS} where the simplest affine
case, i.e. the sine-Gordon theory, has been discussed by Mansfield \cite{PM}.

For the $a^{(1)}_{1}$ Toda field theory, which is the sine-Gordon theory,
the single soliton solution is given by,
\begeq
\phi = \frac{i\sqrt{2}}{\be} \ln
\left(\frac{1-e^{\sig(x-vt)+\rho}}{1+e^{\sig(x-vt)+\rho}}\right),
\endeq
\noi with the constraint $\sig^{2}(1-v^{2})=4m^{2}$. The parameter $\rho =
\eta + i\xi$ is complex, and its imaginary part determines the
topological charge of the soliton. In this case, there are two
possibilities, the soliton and the antisoliton, which have the same mass
equal to $\frac{8m}{\be^2}$. It is well known (see for instance
Rajaraman \cite{Raja})
that beside these soliton solutions, there also exist oscillating
solitonic solutions, or {\it breathers}. The sine-Gordon breather is
constructed from two approaching soliton solutions by changing the velocity
$v$ into $iv$,
\begeq
\phi_{\it breather} = \frac{i\sqrt{2}}{\be}\ln
\left( \frac{1 - e^{\sig(x-ivt)+\rho_{1}} - e^{\sig(x+ivt)+\rho_{2}} -
v^2e^{2\sig x + (\rho_{1} + \rho_{2})}}
{1 + e^{\sig(x-ivt)+\rho_{1}} + e^{\sig(x+ivt)+\rho_{2}} -
v^2e^{2\sig x + (\rho_{1} + \rho_{2})}} \right).
\endeq
\noi Taking $\xi_{1}= -\xi_{2} = -\frac{\pi}{2}$ and
$\eta_{1}=\eta_{2}=\eta$, yields a soliton solution
oscillating about the point $\frac{\ln(v^2)+2\eta}{2\sig}$.
As it is constructed from a soliton-antisoliton pair, this sine-Gordon
breather has zero topological charge; its mass is equal to
$\frac{16m}{\be^{2}\sqrt{1+v^{2}}}$.

Following the prescription of taking {\it imaginary} velocity, the
breathers created from two solitons
of the \an affine Toda theories can be constructed. It turns out that in order
for the energy of the breathers built from two solitons to be real, the
constituent solitons must be of the same mass and approaching each other
with the same imaginary velocity,
resulting in a stationary breather. To obtain a moving breather, one can
apply the usual Lorentz boost to the breather solution. The condition of
real energy also
produces an expression for the masses of these breathers which is less than
the sum of the constituent solitons. One type of breathers can carry
topological charge which coincides with the topological charge of a
certain single soliton, while the other type has zero topological
charge.

This paper is organised as follows. As breather solutions are constructed
{}from soliton solutions, section two discusses the
soliton solutions of the \an series. In section three, the
breathers created from two solitons of
the \an series are determined, and it is shown that there are two types.
Type A breathers are constructed from two solitons of the same species and
type B breathers are constructed from two solitons of opposite species.
The type A breathers carry topological charge while type B
breathers have zero topological charge, these will be discussed in section
four together with some properties of the breather solution. It will
be shown also that these topological charges lie in a fundamental
representation which is a subset of a tensor product representation
of the fundamental representations which are associated to the topological
charges of the constituent solitons. Section five
will discuss the sine-Gordon embedding in some cases of the \an family.
Examples of
the $a_{3}^{(1)}$ and $a_{4}^{(1)}$ breathers are given in section six.
Section seven contains concluding remarks.

\setcounter{equation}{0}

\section{Soliton Solutions}
This section provides a review on the construction of soliton solutions for
the \an affine Toda field theory \cite{TH1,MM}.
Soliton solutions to the affine Toda  equation of motion, (\ref{eom}), can be
derived using Hirota's method. An ansatz for the soliton solutions
of the \an \atft \cite{TH1} is the following,
\begeq\label{phi_soln}
\phi = \frac{i}{\be} \sum_{j=1}^{n} \al_j \ln \frac{\tau_j}{\tau_0}.
\endeq
\noi In Hirota's method \cite{RH}, the $\tau$-functions are a power series
expansion in an arbitrary
parameter $\varepsilon$ which will be set to 1 at the end of the
construction,
\begeq\label{tau}
\tau_{j}=1+\varepsilon t^{(1)}_{j}+\varepsilon^{2} t^{(2)}_{j}+\dots.
\endeq
\noi In the above expression, $t_{j}$ is a function of $(x,t)$.
The equation of motion is
solved order by order in the arbitrary
parameter $\varepsilon$. An $N$ soliton solution is calculated by setting
$t^{(a)}_{j}=0$ for $a > N$.

Inserting the ansatz (\ref{phi_soln}) into the equation of motion
(\ref{eom}), and assuming that the set of $h=n+1$ equations of motion
decouple, results in the following,
\begeq\label{eom_tau}
{\ddot{\tau_{j}}}\tau_{j} - {\dot{\tau_{j}}}^{2} - \tau_{j}''\tau_{j}
+ \tau_{j}'^{2} = m^2(\tau_{j-1}\tau_{j+1} - \tau_{j}^2) \qquad
j=0,1,\ldots,n,
\endeq
\noi where the subscripts of the $\tau$-functions are labelled modulo the
Coxeter number $h=n+1$. A single soliton solution is obtained by keeping the
$\tau$-function expansion up to the order $\varepsilon$ and inserting to the
above equation of motion. This yields,
\begeq\label{1_sol}
\tau^{(a)}_{j} = 1 + \exp[\Omega_{a} + \rho_{a} + ij\theta_{a}]
\endeq
\noi where,
\begeq
\Omega_{a} = \sigma_{a}(x-u_{a}t), \qquad \rho_{a} = \eta_{a} + i\xi_{a},
\qquad \theta_{a} = \frac{2\pi a}{h},
\endeq
\noi $\sigma_a, u_a, \eta_a, \xi_a$ $\in$ $\mbox{\setR}$. The
parameter $\sigma_{a}$ and the velocity $u_{a}$ are related by,
\begeq\label{constraint}
\sigma_a^{2} ( 1 - u_a^2) = 4 m^2 \sin^2 \frac{\pi a }{h}.
\endeq
\noi The superscript $a$ of
the $\tau$-function (\ref{1_sol}) indicates the {\it species} of the
soliton, which can take the value $1 \leq a \leq n$.
One can also write the $\tau$-functions with an explicit dependence on the
{\it lightcone} coordinates,
$x_{\pm} = \frac{1}{\sqrt{2}}(t \pm x)$, then $\Omega_{a}$ becomes,
\[
\Omega_{a} = \delta_{a}^{(-)} x_{+} - \delta_{a}^{(+)} x_{-},
\]
\noi with,
\[
\delta_{a}^{(\pm)} = \frac{1}{\sqrt{2}} \sigma_{a}(1 \pm u_{a}).
\]

A two soliton solution is
obtained by solving the equation of
motion (\ref{eom_tau}) after inserting the expansion of the $\tau$-function
up to second
order in $\varepsilon$. The two soliton solution for {\it species} $a$ and
$b$ is,
\begin{eqnarray}\label{2_sol_a}
\tau^{(ab)}_{j} &=& 1 + \exp[\Omega_{a} + \rho_{a} + ij\theta_{a}] +
\exp[\Omega_{b} + \rho_{b} + ij\theta_{b}] \nonumber \\
& & + A\exp[\Omega_{a} + \Omega_{b} +
\rho_{a} + \rho_{b} + ij(\theta_{a} + \theta_{b})]
\end{eqnarray}
\noi which can be compactly written as,
\begeq\label{2_sol_b}
\tau^{(ab)}_{j} = 1 + (\tau^{(a)}_{j}-1) + (\tau^{(b)}_{j}-1) +
A(\tau^{(a)}_{j}-1)(\tau^{(b)}_{j}-1).
\endeq
\noi The interaction coefficient in the above expression is given by,
\begeq\label{int_A}
A = -
\frac{(\sigma_{a} - \sigma_{b})^2 - (\sigma_{a}u_{a} - \sigma_{b}u_{b})^2 - 4
m^2 \sin^2( \frac{\pi}{h} (a-b))}{(\sigma_{a} + \sigma_{b})^2 - (\sigma_{a}
u_{a} + \sigma_{b} u_{b})^2 - 4 m^2 \sin^2( \frac{\pi}{h} (a+b))}.
\endeq
\noi If the rapidity variable $\Theta_a$ is defined through the
velocity $u_a$
as,
$u_{a} = \tanh \Theta_{a}$, then the interaction coefficient can be written
in terms of rapidity difference $\Theta= \Theta_{a} -
\Theta_{b}$ as,
\begeq\label{int_A_rap}
A = \frac{\sin\left(\frac{\Theta}{2i} +
\frac{\pi(a-b)}{2h}\right) \sin\left(\frac{\Theta}{2i} -
\frac{\pi(a-b)}{2h}\right)} {\sin\left(\frac{\Theta}{2i} +
\frac{\pi(a+b)}{2h}\right) \sin\left(\frac{\Theta}{2i} -
\frac{\pi(a+b)}{2h}\right)}.
\endeq

As in the case for two soliton solutions, a
multi-soliton solution can be constructed from a collection of single
soliton solutions \cite{TH1}. The interaction between the solitons is
pair-wise with the interaction coefficient as given above.

The energy-momentum tensor of
soliton solutions
can be written as \cite{OTU1},
\begeq\label{en_mom_tensor}
T_{\mu \nu} = (\eta_{\mu \nu}\partial^{2} - \partial_{\mu}\partial_{\nu})
C.
\endeq
\noi Or in terms of the lightcone components,
\begin{eqnarray}
\label{1_diff_eq}
T_{+-} &=& \partial_{+}\partial_{-}C ,\\
\label{2_diff_eq}
T_{\pm\pm} &=& - \partial_{\pm}^{2}C.
\end{eqnarray}
\noi $T_{+-}$ is actually the trace of the energy-momentum tensor, then
(\ref{1_diff_eq}) becomes,
\begeq\label{diff_eq_C}
\partial^{2}C = -\frac{2m^{2}}{\be^2} \sum_{j=0}^{n}
\left(e^{i\be\al_{j}\cdot\phi} - 1 \right).
\endeq
\noi Using the soliton solutions in the energy-momentum tensor leads to
the following solution for the function $C$ up to a constant,
\begeq\label{C}
C = -\frac{2}{\be^{2}} \sum_{j=0}^{n} \ln \tau_{j}.
\endeq

The mass of the soliton solution can be calculated as follows.
The energy and momentum density is given in terms of components of the
energy-momentum density as, ${\cal E} = T_{00}$ and ${\cal P} = T_{10}$.
Consider instead, the {\it lightcone} energy-momentum density,
\begeq\label{Plightcone1}
{\cal P^{\pm}} = \frac{{\cal E \pm P}}{\sqrt{2}},
\endeq
\noi Thus, using (\ref{en_mom_tensor}), the components of the lightcone
energy-momentum are given by,
\begeq\label{Plightcone2}
P^{+} = (-\partial_{+}C)_{x=-\infty}^{\infty}, \qquad \qquad
P^{-} = (\partial_{-}C)_{x=-\infty}^{\infty}.
\endeq
\noi For the single soliton solution (\ref{1_sol}), one finds that
\begeq
\partial_{\pm}C = \mp \frac{2}{\be^2} \sum_{j=0}^{n}
\frac{\de_{a}^{\mp}(\tau_{j}^{(a)}-1)}{\tau_{j}^{(a)}},
\endeq
\noi In the limit as $x \rightarrow -\infty$ the ratios
$\frac{(\tau_{j}^{(a)}-1)}{\tau_{j}^{(a)}}$ vanish, and in the limit $x
\rightarrow \infty$ all the ratios tend to 1. Recall the constraint
(\ref{constraint}) and write the rapidity of the soliton as $\Theta_{a}
= \frac{1}{2}\ln\left(\frac{1+u_{a}}{1-u_{a}}\right)$, then the
lightcone energy-momentum of a single soliton is given by
\begeq
P^{\pm} = \frac{4hm}{\sqrt{2}\be^{2}}\sin\left(\frac{\theta_{a}}{2}\right)
e^{\mp\Theta_{a}}
\endeq
\noi And its mass is calculated to be
\begeq\label{1_sol_mass}
M_{a}^{2} = 2P^{+}P^{-} =
\left(\frac{4hm}{\be^{2}}\sin\left(\frac{\theta_{a}}{2}\right)\right)^{2}.
\endeq
\noi Note, that the mass
of the species $a$ soliton is
proportional to the mass of the fundamental particle of the \an affine
Toda field theory, $m_{a} = 2m\sin(\frac{\theta_{a}}{2})$.

For a multi-soliton solution the calculation above can be performed in
the same way. In particular a two soliton solution of species $a$ and
$b$ yields,
\begeq\label{en_2_sol}
\sqrt{2}P^{\pm} = M_{a}e^{\mp \Theta_{a}} + M_{b}e^{\mp \Theta_{b}}.
\endeq

The single solitons of species $a$ carry topological charges which lie in the
highest weight representation of the $a^{th}$ fundamental weight of the
$A_{n}$ Lie algebra \cite{WAM}. Thus, it
is natural to associate the species of the solitons with the nodes of the
Dynkin diagram of the associated Lie algebra, $A_{n}$.

\setcounter{equation}{0}
\section{Breather Solutions}
To obtain an oscillating solitonic solution which is constructed from two
solitons, one can follow the sine-Gordon prescription of changing the
velocity into an imaginary velocity, i.e. changing $u$ into $iv$ in the
$\tau$-functions.
Care must be taken in the analytic continuation of $u \rightarrow iv$ such
that the energy and
momentum are real, although the energy and momentum densities
generally are complex.

Changing $u$ into $iv$ also means changing a real rapidity into an
imaginary rapidity, with a relation between velocity $v$ and rapidity
$\Theta$ becomes $v=\tan(-i\Theta)$. From the lightcone
energy-momentum of the two
soliton solution (\ref{en_2_sol}), a real energy and momentum can be
achieved
provided that the two solitons forming a breather are of the same mass and
moving towards to each other with the same velocity giving a {\it
stationary} breather.
One can make an oscillating solution from solitons of two different
masses, but the energy and momentum of
this solution will not be real. Generally, one can add
a real rapidity $\Theta_{0}$ as a phase in the energy-momentum tensor, which
acts as a Lorentz boost to the breather solution. Thus,
\begeq\label{en_bre_gen}
P^{\pm}_{breather} =
\frac{4hm_{a}}{\sqrt{2}\be^2}\cos(\Theta_{a})e^{\mp\Theta_{0}},
\endeq
\noi $m_{a}$ above is the mass of the fundamental particles of the \an
affine Toda theory. For simplicity, in what follows
only stationary breathers are considered. Hence (\ref{en_bre_gen}) becomes,
\begeq\label{en_bre}
P^{\pm}_{breather} = \frac{4h}{\sqrt{2}\be^2} \frac{m_{a}}{\sqrt{1+v^{2}}},
\endeq
\noi and the mass of a breather is,
\begeq\label{bre_mass}
M_{breather} = \frac{2M_{a}}{\sqrt{1+v^{2}}} =
\frac{4hm_{a}}{\be^2\sqrt{1+v^{2}}}. \endeq
\noi It is obvious that the mass of a breather is less than the sum of its
constituent solitons. This result generalizes the sine-Gordon case,
i.e. taking $h=2$ gives the mass of the sine-Gordon breather.

The masses $m_{a}$ of the fundamental particles of \an series are
degenerate with respect to the $Z_{2}$ symmetry of the $A_{n}$ Dynkin
diagram, i.e.  $m_{a} = m_{h-a}$. Hence,
there are two possibilities of forming a breather. Either the two
constituent solitons are of the same species, these breathers will
be called {\it type} A breathers or, the two constituent solitons are of
opposite species, {\it type} B breathers. Exceptions to this
classification are the breathers constructed from solitons of species
$(n+1)$ of the $a_{2n+1}^{(1)}$ theories. These breathers are
sine-Gordon embedded breathers which belong to
both type A and B as will be explained in the following sections.

Looking back at the $\tau$-function of a two soliton solution of the same
constituent mass, choosing
$u_{a}=-u_{b}=iv$ yields the breather $\tau$-function,
\begin{eqnarray}\label{tau_bre1}
\tau^{(ab)}_{j} &=& 1 + \exp[\sigma_{a}(x-ivt) + \rho_{a} + ij\theta_{a}] +
\exp[\sigma_{b}(x+ivt) + \rho_{b} + ij\theta_{b}] \nonumber \\
& & + \exp[\sigma_{+}x + \lambda + \rho_{+} + ij\theta_{+}],
\end{eqnarray}
\noi the interaction coefficient is written as $A=e^{\lambda}$ with
$\lambda = \zeta + i\delta$, where $\zeta,\delta$ $\in$ $\setR$ and
\[
\sigma_{+} = \sigma_{a} + \sigma_{b}, \qquad
\rho_{+} = \rho_{a} + \rho_{b}, \qquad
\theta_{+} = \theta_{a} + \theta_{b},
\]
\[
\eta_{-} = \eta_{a} - \eta_{b}, \qquad
\xi_{+} = \xi_{a} + \xi_{b}.
\]
\noi Note that for solitons of the same mass, $\sig_{a}=\sig_{b}$.
By the ansatz (\ref{phi_soln}), it is clear that in order to have a
well-defined solution, each $\al_{j}$ component of the solution
$\phi$ must be well defined. Thus, for each $j$,
the ratio $\frac{\tau_{j}}{\tau_{0}}$ must not become zero or
infinite.
Evaluation of the behaviour of the $\tau$-function can be done easily
by writing the real
and imaginary part of (\ref{tau_bre1}) explicitly. It turns out that to
avoid the real and imaginary part of (\ref{tau_bre1}) becoming zero
simultaneously at the same point, the parameters $\xi_{+}$ and
$\eta_{-}$ are restricted to a certain range of definition.

Moreover, for type A and B breathers, the interaction coefficient can have
either positive or negative value. The critical velocity when
the interaction coefficient changes sign is, for a type A breather
\begeq\label{vc_A}
v^{(A)}_{c} = \tan(\frac{\theta_{a}}{2}),
\endeq
\noi and for a type B breather,
\begeq\label{vc_B}
v^{(B)}_{c} = \frac{1}{v^{(A)}_{c}}.
\endeq
\noi For breathers with constituent solitons of species $a=\frac{h}{2}$,
the interaction coefficient never changes sign.

%% type A
%%%%%%%%%%%%%%%%%%%%%%%%%%%%%%%%%%%%%%%%%%%%%%%%%%%%%%%%%%%%%%%%%%%%%%%
\subsection{Type A Breathers}
The breathers with constituent solitons of the same
species will have a negative
interaction coefficient, $A < 0$,
when $v^{2} < v^{(A)2}_{c}$ (see figure 1). For these breathers, both
the real and imaginary part of (\ref{tau_bre1}) will be trivially zero
when the parameters $\xi_{+}$ have the following values,
\begeq\label{av-}
\xi_{+} = (2\pi - j\theta_{+}) \mbox{ mod } 2\pi, \qquad j=0,\ldots,n.
\endeq
\noi This divides the range of $\xi_{+}$ into several regions
which will determine the topological charge of the breather.

Furthermore, the maximum ``distance'' of separation between the
two constituent solitons is restricted as follows,
\[
-\min_{j}(\eta^{j}_{c}) < \eta_{-} < \min_{j}(\eta^{j}_{c}),
\]
\noi where,
\[
\eta^{j}_{c} = \mbox{arcosh}\left[2|A|\sin^{2}\frac{1}{2}(\xi_{+} +
j\theta_{+}) + 1\right],
\]
\noi with $j=0,\ldots,n$. Each $\eta^{j}_{c}$ above will restrict
$\tau_{j}$ such that it will never be zero. Hence, for all
$\tau$-functions to
avoid zero, the minimum value of $\eta^{j}_{c}$ is taken as the limit on
the range of $\eta_{-}$.

For breathers of type A with positive interaction coefficient, the
parameters
$\xi_{+}$ must not have the following values,
\begeq\label{av+}
\xi_{+} = (\pi - j\theta_{+}) \mbox{ mod } 2\pi, \qquad j=0,\ldots,n.
\endeq
\noi The separation ``distance'' is restricted as above with,
\[
\eta^{j}_{c} = \mbox{arcosh}\left[2A\cos^{2}\frac{1}{2}(\xi_{+} +
j\theta_{+}) - 1\right].
\]
\noi As $\eta^{j}_{c}$ are defined through an arcosh-function, these
$\eta^{j}_{c}$ in turn will restrict the allowed velocity of
the constituent solitons. Generally, not all $v^{2} > v^{(A)2}_{c}$ are
allowed. In fact all velocities with absolute value greater than the
absolute value of the critical velocity are allowed if for all $j$ the
following is true,
\[
\left[1-\frac{\cos^{2}\frac{1}{2}(\xi_{+} +
j\theta_{+})}{\cos^{2}(\frac{\theta_{+}}{4})}\right] < 0.
\]
\noi Otherwise, the velocity
is bounded from above by,
\[
v^{(A)2}_{c} < v^{2} \leq
\frac{v^{(A)2}_{c}}{{\displaystyle
\max_{j}}\left[1-\frac{\cos^{2}\frac{1}{2}(\xi_{+}
+ j\theta_{+})}{\cos^{2}(\frac{\theta_{+}}{4})}\right]}.
\]

%% type B
%%%%%%%%%%%%%%%%%%%%%%%%%%%%%%%%%%%%%%%%%%%%%%%%%%%%%%%%%%%%%%%%%%%%%%%
\subsection{Type B Breathers}
The constituent solitons of type B breathers are of opposite species. The
negative interaction coefficient regime is accomplished with $v^{2} >
v^{(B)2}_{c}$. In this case, the real and imaginary part of the
$\tau$-functions will be trivially zero for
\begeq\label{bv}
\xi_{+} = 0 \mbox{ mod } 2\pi,
\endeq
\noi i.e. the parameter $\xi_{+}$ must not be an integer multiple of
$2\pi$. For each $\tau_{j}$ to avoid zero, the separation parameter
$\eta_{-}$ is limited to take values between,
\[
-\min_{j}(\eta^{j}_{c}) < \eta_{-} < \min_{j}(\eta^{j}_{c}),
\]
\noi where,
\[
\eta^{j}_{c} = \mbox{arcosh}\left[2|A|\sin^{2}\frac{1}{2}(\xi_{+} +
j\theta_{+}) + 1\right],
\]
\noi with $j=0,\ldots,n$. As for type A breathers, each $\tau_{j}$ has
its own $\eta^{j}_{c}$, and the smallest of these
is taken as the limit.

Contrary to the type A breathers, it is not possible to have type B
breathers with
positive interaction coefficient, or to have velocity $v^{2} <
v^{(B)2}_{c}$.
Since the $\tau$-functions would necessarily pass the origin and this
would lead to a solution which is not well-defined.

\setcounter{equation}{0}
\section{Properties of the Breather Solutions}

%% interaction coefficient
%%%%%%%%%%%%%%%%%%%%%%%%%%%%%%%%%%%%%%%%%%%%%%%%%%%%%%%%%%%%%%%%%%%%%%%%
\subsection{The Interaction Coefficient}
The interaction coefficient A has properties similar to the properties of
the
$S$-matrix of the fundamental Toda particles. Recall the expressions for
A from (\ref{int_A}) and (\ref{int_A_rap}). The interaction term of
the type A breathers is,
\begeq
A_{aa} = \frac{v^{2}}{(1+v^{2})\cos^{2}(\frac{\theta_{a}}{2})-1},
\endeq
\noi or in terms of rapidity difference $i\Theta$,
\begeq
A_{aa} = \frac{\sin^{2}(\frac{i\Theta}{2})}{\sin(\frac{i\Theta}{2}
+ \frac{\theta_{a}}{2})\sin(\frac{i\Theta}{2} - \frac{\theta_{a}}{2})}.
\endeq
\noi And for type B breathers,
\begeq
A_{a{\bar a}} = (1+v^{2})\cos^{2}(\frac{\theta_{a}}{2})-v^{2},
\endeq
\noi in terms of rapidity difference $i\Theta$,
\begeq
A_{a{\bar a}} = \frac{\cos(\frac{i\Theta}{2} +
\frac{\theta_{a}}{2})\cos(\frac{i\Theta}{2} -
\frac{\theta_{a}}{2})}{\cos^{2}(\frac{i\Theta}{2})}.
\endeq

As the case of $S$-matrices of fundamental Toda particles, these
interaction coefficients admit a pole.
For type A breathers,
\begeq
v = v^{(A)}_{c} \qquad \mbox{ or, } \qquad i\Theta = \frac{2\pi a}{h}
\endeq
\noi and, for type B breathers,
\begeq
v \rightarrow \infty \qquad \mbox{ or, } \qquad i\Theta = \pi.
\endeq
\noi It is readily seen that the pole of $A_{aa}$ is exactly the fusing
angle related to the process $a + a \rightarrow {\overline{(h-2a)}}$ of the
fundamental particles \cite{BCDS1}. Hollowood noted that the
same fusing rule also applies to soliton fusings in
$a_{n}^{(1)}$ theories \cite{TH1}. In
fact, the fusing rule of fundamental particles applies also to all {\it
simply laced} affine Toda solitons \cite{OTU1,RAH}.
The fusing of two solitons of species $a$ into ${\overline{(h-2a)}}$ hinted
that the topological charge of type A breathers has to be found in the same
representation as the topological charges of ${\overline{(h-2a)}}$ single
solitons.

It is not surprising that at the pole of the interaction coefficient, the
breathers fail to exist.
Using the breather $\tau$-function, (\ref{tau_bre1}), with the interaction
coefficient approaching its pole, the interaction term dominates. Hence,
the solution falls into one of the vacuum solutions of the complex
affine Toda potential,
\begeq
\phi = -\frac{1}{\be} \sum^{n}_{j=1} j\al_{j}\theta_{+}.
\endeq
\noi On the other hand, if the positions of the constituent solitons are
simultaneously shifted by $-\frac{\zeta}{2}$, i.e. changing $\eta
\rightarrow \eta - \frac{\zeta}{2}$, the breather turns into
a static solution as the interaction coefficient approaches its pole.

When there is no interaction between the constituent solitons, the
interaction coefficient is zero. In this case, the $\tau$-functions
do not give a well defined solution, as
the $\tau$-functions will vanish at a particular point in space-time.
This is
quite obvious, since the breathers are bound states of soliton pairs and
therefore a breather interaction coefficient cannot be zero.

Furthermore, the interaction coefficient $A$ has the following
general properties some of which are similar but not the same as the
properties of the $S$-matrix,
\begin{itemize}
  \item {\it Crossing symmetry}
        \begeq
        A_{aa}(v) = A^{-1}_{a{\bar a}}(\frac{1}{v}) \qquad \mbox{ or, }
        \qquad A_{aa}(i\Theta) = A^{-1}_{a{\bar a}}(i\Theta - \pi).
        \endeq
  \item {\it Evenness}
        \begeq
        A(v) = A(-v) \qquad \mbox{ or, } \qquad A(i\Theta) = A(-i\Theta).
        \endeq
  \item {\it Symmetry}
        \begeq
        A_{a{\bar a}}(v) = A_{{\bar a}a}(v) \qquad \mbox{ or, }
        \qquad A_{a{\bar a}}(i\Theta) = A_{{\bar a}a}(i\Theta).
        \endeq
  \item { \it Periodicity}
        \begeq
        A(i\Theta) = A(i\Theta + 2\pi).
        \endeq
\end{itemize}

%% topological charges
%%%%%%%%%%%%%%%%%%%%%%%%%%%%%%%%%%%%%%%%%%%%%%%%%%%%%%%%%%%%%%%%%%%%%
\subsection{The Topological Charges}
In the soliton solutions of the complex affine Toda field equations, the
topological charge is a conserved quantity of zero spin. For the \an
series, topological charges of the single and
multi-soliton solutions have been calculated
\cite{WAM}. Unfortunately, the calculation
of topological charges of multi-soliton solutions used there is
not applicable for breathers because the two solitons constituting the
breather only separate a finite ``distance'' from each other.

The topological charge $q$ of a solution $\phi$ is defined by
\begin{equation}\label{t1}
q = \frac{\be}{2\pi} \int_{-\infty}^{\infty} \partial_x \phi \mbox{d}x =
\frac{\be}{2\pi}
\lim_{x \to \infty} (\phi(x,t) - \phi(-x,t)).
\end{equation}
\noi Writing the solution $\phi$ in terms of the breather
$\tau$-functions, defining $f_j = \frac{\tau_j}{\tau_0} $ for $ j=1
\dots n$ and making use of the definition of the logarithm of a
complex number, (\ref{t1}) can be recast as
\begin{eqnarray}\label{t2}
q = -\frac{1}{2\pi i} \sum_{j=1}^{n} &\alpha_j& \; \lim_{x \to \infty}
\{\ln |f_j(x,t)| - \ln |f_j(-x,t)| \nonumber\\
&+& i \arg(f_j(x,t)) + 2 i \pi k' - i
\arg(f_j(-x,t)) -2 i \pi k'' \},
\end{eqnarray}
\noi where $k',k'' \in \setZ$.
A simplification of (\ref{t2}) results from the fact that $\lim_{|x| \to
\infty }|f_j(x,t)| = 1 $, thus
\begin{eqnarray}\label{t3}
q= -\frac{1}{2\pi} \sum_{j=1}^{n} \alpha_j \; \lim_{x \to \infty}
\{\arg(f_j(x,t)) -\arg(f_j(-x,t)) + 2 \pi k \}
\end{eqnarray}
\noi with $k= k'-k''$.
The number $k$ determines the curve $f_j$ in the
complex plane, and in particular, how often and in what direction it winds
around the origin.
The topological charge is therefore determined by the change in the
argument of $f_j$ as $|x|$ goes to infinity.

%type A starts here
%%%%%%%%%%%%%%%%%%%%%%%%%%%%%%%%%%%%%%%%%%%%%%%%%%%%%%%%%%%%%%%%%%%%%
For the type A breather it is not
too difficult to deduce the number of distinct topological charges in the
fashion of \cite{WAM}. The number of the topological charges is
determined by the number of sectors of allowed values of $\xi_{+}$.
{}From the expression for forbidden $\xi_{+}$'s, (\ref{av-}) or (\ref{av+}),
it follows that one looks for the smallest number $p$ for which
\begin{equation}\label{top_num}
\frac{2ap}{h} = k \, \mbox{with} \, p,k \in \setN .
\end{equation}
Here $a$ is the species of the constituent solitons and $h$ is the
Coxeter number.

With $2\tilde a = \frac{2a}{\gcd(2a,h)}$, $\tilde h =
\frac{h}{\gcd(2a,h)}$ then (\ref{top_num}) can be rewritten as
\begin{equation}
2\tilde ap = \tilde h k .
\end{equation}
Because $2\tilde a$ and $\tilde h$ are coprime it follows that
$p=\tilde h$ and $k = 2\tilde a$. Thus the range of the allowed values for
$\xi_{+}$ is divided into $\tilde h$ sectors. This leads to the following
formula for the maximum number of topological charges of type A breather
with constituent solitons of species $a$
\begin{equation} \label{tn}
\tilde h = \frac{h}{\gcd(2a,h)}.
\end{equation}
\noi This argument holds independently of the sign of the interaction
coefficient.
The argument of $f_j(x,t)$ can only change when $f_j(x,t)$ is
undefined or zero, hence the topological charge within each sector is
constant. It turns out that in each of these
sectors,
the topological charges take a different unique value. These topological
charges are related by permutation of the roots $\al_{j}$ for $j=0,\ldots,n$.
The topological charge of a specific sector will
be determined first, and is called the {\it highest charge}
\cite{WAM}. Then it
will be shown that in all other sectors, the topological charges will be
different. This means that $\tilde h$ is indeed the number of topological
charges associated with a breather.

In the following, calculation of the highest charge will be performed for
a type A breather with negative interaction coefficient. The calculation
for positive interaction coefficient is the same and will not be presented
here. To calculate the highest topological charge, one has to employ a
little trick first to simplify the breather $\tau$-function.
The type A breather $\tau$-function is given by
\begin{eqnarray*}
\tau_{j}^{(aa)} &=& 1 + \exp[\sigma_{a}(x+ivt)+\rho+ij\theta_{a}] +
\exp[\sigma_{a}(x-ivt)+\rho'+ij\theta_{a}] \\
&& -\exp[\zeta + 2\sigma_{a}x+\rho_+ +2ij\theta_{a}].
\end{eqnarray*}
\noi First choose $t=0$ because the topological charge does not depend on
the time. Second choose $\rho =
{-\zeta /2} + \hat \rho$ and
$\rho' = {-\zeta /2} + \hat \rho'$, this corresponds to a
simultaneous shift of the constituent solitons to the left. By this
shifting, the last term in the breather $\tau$-function will not
depend on the interaction coefficient $A$,
\[
\tau_j^{(aa)} = 1 + \exp(\sigma_{a}x+ij\theta_{a} - \zeta /2 ) (e^{\hat
\rho} + e^{\hat \rho'}) -\exp( 2\sigma_{a}x+\hat \rho_+ + 2ij\theta_{a}).
\]
\noi With $\mu_a(2j) = \frac{4\pi a j}{h} \mbox{ mod } 2 \pi$,
the limits $|x| \to \infty$ of $f_j$ will give
\begin{eqnarray}\label{tcd}
\lim_{x \to \infty} f_j &=& e^{i\mu_a(2j)}\nonumber ,\\
\lim_{x \to - \infty}f_j &=& 1.
\end{eqnarray}
\noi Moreover one
can take the limit $\zeta$ approaching $+\infty$, this
corresponds to
choosing the velocity $v$ very near to $v^{(A)}_{c}$. As long as $v$ is
not equal to $v^{(A)}_{c}$ the breather solution is
well-defined by construction. Write $y = e^{2\sigma_{a} x}$, then provided
one
does not take the limit $x \to \infty$, the $y^{1/2}$ term can be dropped,
\begin{eqnarray}\label{simplified_tau}
\tau_j^{(aa)} &=& 1 + y^{1/2} \exp(ij\theta_{a} - \zeta /2 ) (e^{\hat \rho} +
e^{\hat \rho'}) - y \exp(\hat \rho_+ +2ij\theta_{a}) \nonumber \\
&=& 1 - y \exp(\hat \rho_+ +i\mu_a(2j)).
\end{eqnarray}

By splitting the ratio $f_j$ into its real and imaginary part one can
now easily show that $f_j$ traces out a clockwise curve in the
complex plane, i.e. the winding number $k$ is zero.
To see this take $\hat \rho = \hat \rho' = i (\pi - \frac{\varepsilon}{2})$
where $\varepsilon $ is a real, positive and infinitesimal parameter,
\[
\tau_j= 1 - y \exp(i(\mu_a(2j) - \varepsilon)).
\]
\noi Then the ratio $f_j$ can be written as
\begin{eqnarray*}
f_j &=&\frac{1}{|1 - y e^{\hat \rho_{+}}|^2}
[1 - y(\cos (\mu_a(2j) - \varepsilon) + \cos(\varepsilon) ) + y^2 \cos
(\mu_a(2j)) \\
&&+i\{- y (\sin (\mu_a(2j) - \varepsilon) + \sin (\varepsilon) ) + y^2 \sin
(\mu_a(2j))\}].
\end{eqnarray*}
\noi The only zeros for the imaginary part occur when $y=0$ and
\[
y = \frac{\sin(\mu_a(2j) - \varepsilon) +
\sin(\varepsilon)}{\sin(\mu_a(2j))}, \]
\noi with $\mu_a(2j) \neq 0 \mbox{ or } \pi$. For small $\varepsilon$ this is
\begeq\label{im_y}
y = 1 + \varepsilon \frac{1-\cos(\mu_a(2j))}{\sin(\mu_a(2j))} +
O(\varepsilon^2).
\endeq
\noi Now inserting (\ref{im_y}) into the real part of $f_j$ results in,
\[
\Re e (f_j |1 - y e^{\hat \rho_{+}}|^2) = -2 \varepsilon
\frac{1-\cos(\mu_a(2j))}{\sin(\mu_a(2j))} + O(\varepsilon^2).
\]
\noi One should also observe that in the small $\varepsilon$ regime the
imaginary part behaves for small $y$ like
\[
\Im m (f_j |1 - y e^{\hat \rho_{+}}|^2) = - y \sin (\mu_a(2j)) + \dots .
\]
\noi So, for $0 < \mu_a(2j) < \pi$, the curve starts at $(1,0)$ with a
negative
imaginary part and crosses the negative part of the real line. For
$\pi < \mu_a(2j) < 2\pi$, it starts at $(1,0)$ with a positive imaginary
part and crosses the positive part of the real line. In any case, it
winds around the origin in the clockwise sense. Thus, the change of
argument of $f_j$ is given by $\mu_{a}(2j)-2\pi$.
The explicit formula for the {\it highest topological charge} is therefore
determined by (\ref{t3})
\begeq\label{high_q}
q_{a}^{(1)}= \sum_{j=0}^{n} \frac{2a(h-j)\mbox{ mod } h}{h} \al_{j}.
\endeq
\noi In the summation above, the extended root $\al_{0}$ is included for
convenience in the permutation of the simple roots and $\al_{0}$.
As mentioned previously, this result does not depend on the sign of the
interaction coefficient.

{}From this highest charge, one can obtain all the other charges as follows.
Suppose initially the value of $\xi_{+}$ is chosen.
Then, making a shift of $\frac{4\pi a}{h}$ on this $\xi_{+}$ amounts to
sending the breather solution to a different sector of $\xi_{+}$.
Successive applications of this shift will bring the breather solution to
every allowed sector of $\xi_{+}$. With the $\tilde h^{th}$
application it will return to the original sector.
Recall that with (\ref{simplified_tau}) the breather solution is given
by,
\[
\phi = \frac{i}{\be} \sum_{j=0}^{n} \al_{j}
\ln(1-\omega_{a}^{2j}ye^{\hat \rho_{+}}).
\]
\noi Making the shift $\xi_{+} \to \xi_{+} - \frac{4 \pi a}{h}$ in the above
solution gives,
\[
\phi = \frac{i}{\be} \sum_{j=0}^{n} \al_{j}
\ln(1-\omega_{a}^{2j}ye^{\hat \rho_{+} - \frac{4i \pi a}{h}}) =
\frac{i}{\be} \sum_{j=0}^{n} \al_{j+1}
\ln(1-\omega_{a}^{2j}ye^{\hat \rho_{+}}).
\]
\noi Thus this shifting
is the same as cyclically permuting the roots $\al_{j}$ for all
$j=0,\ldots,n$. And hence, each shifting results to a different
topological charge. Since the maximum number one can shift $\xi_{+}$ is
$\tilde h$ times, then $\tilde h$ is exactly the number of topological
charges of the breather solution. The expression for all the topological
charges is
\begeq\label{q_k}
q_{a}^{(k)} = \sum_{j=0}^{n} \frac{2a(h-j)\mbox{ mod }h}{h}
\al_{(j+k-1)}, \qquad k=1,2,\ldots,{\tilde h},
\endeq
\noi where the roots $\al_{j}$ are labelled modulo $h$.

This is analogous to the one soliton case \cite{WAM}.
Furthermore, as explained in the same paper, all these topological
charges lie in the same representation because they are related by a
Weyl transformation as will be shown in the next subsection.

%type A ends here
%%%%%%%%%%%%%%%%%%%%%%%%%%%%%%%%%%%%%%%%%%%%%%%%%%%%%%%%%%%%%%%%%%%%%%%

%type B starts here
%%%%%%%%%%%%%%%%%%%%%%%%%%%%%%%%%%%%%%%%%%%%%%%%%%%%%%%%%%%%%%%%%%%%%
For the type B breather it has been determined in a preceding
caculation (subsect. 3.2) that there is only one sector of allowed values
for $\xi_+$. The only possible way for the
topological charge to change is whenever the ratio $f_j$ is not
well-defined, i.e $\xi_+$  changes from one sector to another. So, in
this case there cannot be a change in the topological charge. The only
open question now is what value the topological charge takes. To
determine this one simply follows the previous prescription
\cite{WAM}. The $\tau$-functions for the type B breather are given by
\begin{eqnarray*}
\tau_j^{(a{\bar a})} &=& 1 +
\exp[\sigma_{a}(x+ivt)+\rho+ij\theta_{a}] +
\exp[\sigma_{a}(x-ivt)+\rho'-ij\theta_{a}] \\
&& -\exp[\zeta + 2\sigma_{a}x+\rho_+],
\end{eqnarray*}
\noi where $\bar a = h-a$.
Because the topological charge is time independent
one can set $t=0$. Also, one can substitute $\exp(\sigma_a x) =z$,
$\rho = \rho' = i(\pi + \frac{\epsilon}{2})$ with $ \epsilon \in
\setR$ and infinitesimal.
The $\tau$-function is then given in the compact form
\[
\tau_j^{(a{\bar a})} = 1 - 2 z \cos (j\theta_{a}) e^{i\frac{\epsilon}{2}}-
z^2 e^{\zeta + i \epsilon}.
\]
\noi Let $f_{j}$ be defined as before.
The start and
end point of the curve traced out by $f_j$ as $x$ goes from $-\infty$ to
$\infty$  are in this
case the same, $f_{j}(x=\pm\infty)=1$. Solving an equation for the imaginary
part of $f_j$ one
finds that these are also the only points for which the imaginary
part vanishes. Therefore the winding number $k$ is zero, because
the curve cannot wrap around the origin. Moreover, since the
change of arguments of $f_j$ as $x$ goes from $-\infty$ to
$\infty$ is zero, the topological charge of any type B breather is
deduced to be zero.
In a sense, type B breathers are {\it sine-Gordon like} breathers. The
constituent solitons are of opposite topological charges such that the
resulting breather has zero topological charge. In fact, as will be
discussed
in the next section, type B breathers do not come from a sine-Gordon
embedding in the theory.

%type B ends here
%%%%%%%%%%%%%%%%%%%%%%%%%%%%%%%%%%%%%%%%%%%%%%%%%%%%%%%%%%%%%%%%%%%%%
%% topological charge and rep.
%%%%%%%%%%%%%%%%%%%%%%%%%%%%%%%%%%%%%%%%%%%%%%%%%%%%%%%%%%%%%%%%%%%%%
\subsection{Topological Charge and Representation Space}
It is natural to expect that the topological charges which have been
derived in the previous calculation lie in the tensor product
representation of the fundamental representation associated with the
topological charges of the constituent solitons. In fact,
for the type A breather, with the exception for breathers built from
species $(n+1)$ in the $a_{2n+1}^{(1)}$ cases, the topological charges
lie in the fundamental representation which is a component of the
Clebsch-Gordan decomposition of the tensor product representation.
For type B breathers and the exceptional
cases above, the topological charge (which is zero) lies in the singlet
representation component of the Clebsch-Gordan decomposition of the
tensor product representation.

For the non-zero highest topological charge, the first step is to show
that it lies
in the ${\cal R}_{\lambda_{2a \bmod h}}$ fundamental representation.
This will be shown using a combination of Weyl transformations
\cite{WAM}. Then the
second step is to show the other topological charges are related to the
highest charge by a special Coxeter element of the Weyl group.
It is convenient to write the highest charge (\ref{high_q}) as
\begeq\label{high_q2}
q_{a}^{(1)}= \sum_{j=0}^{n} \frac{bj\mbox{ mod } h}{h} \al_{j},
\endeq
\noi where $b=h - (2a \bmod h)$. Because of the $Z_{2}$ symmetry of
the simple roots, it is necessary to consider only the case $b \leq
\left[\frac{h}{2}\right]$. The notation $[x]$ means the largest integer
less or equal to $x$, hence for $h$ even,
$\left[\frac{h}{2}\right]=\frac{h}{2}$, and for $h$ odd,
$\left[\frac{h}{2}\right]=\frac{h-1}{2}$.
Furthermore, (\ref{high_q2}) can be rewritten in terms of the fundamental
weights $\lambda_j$ defined by
$\frac{2\lambda_{j}\cdot\al_{k}}{\al_{k}^{2}}=\delta_{jk}$ as follows,
\begin{eqnarray}\label{high_q3}
q_{a}^{(1)} &=& \frac{1}{h} \{2[b \bmod h] - [2b \bmod h]\} \lambda_{1} +
\frac{1}{h}
\{2[bn \bmod h] - [b(n-1) \bmod h]\} \lambda_{n} \nonumber \\
&& + \sum_{j=1}^{n-1} \frac{1}{h}\{2[bj \bmod h] -
[b(j-1) \bmod h] - [b(j+1) \bmod h]\} \lambda_{j}.
\end{eqnarray}
\noi Then the following can be demonstrated easily,
\begin{eqnarray}
q_{a}^{(1)} \cdot \al_{j} = \left\{
\begin{array}{lcl}
  1 && j = n, \\
  0 \mbox{ or } -1 && j = n-1, \\
  0 \mbox{ or } -1 \mbox{ or } 1 && 1 \leq j < n-1.
\end{array}
\right.
\end{eqnarray}
The part $q_{a}^{(1)} \cdot \al_{j} = -1$ for $j < n-1$ will be
demonstrated in the following. Let, $bj = ch + d$ where $d<b$ and $c \geq
0$, thus $j=1$ is excluded. Then with (\ref{high_q3}) one find that,
\begin{eqnarray*}
q_{a}^{(1)} \cdot \al_{j} &=& \frac{1}{h}\{2[bj \bmod h] - [b(j-1) \bmod
h] - [b(j+1) \bmod h] \} \\
&=& \frac{1}{h}\{2d - (d-b+h) -(d+b)\} = -1.
\end{eqnarray*}
\noi There are $(b-1)$ terms of $q_{a}^{(1)} \cdot \al_{j} =
-1$ for $j < n-1$ since this happens only when $d<b$.
Furthermore, it is straightforward to see that for $1<j<n-1$
\begeq
q_{a}^{(1)} \cdot \al_{j} = -1 \Longrightarrow q_{a}^{(1)} \cdot
\al_{j-1} = 1.
\endeq

Thus, if the scalar products of $q_{a}^{(1)}$ with the simple roots
$\{\al_{j}\}$ is written as a row vector, it has the entry 1 at $n^{th}$
position and there are $(b-1)$ pairs of $(1,-1)$ to the left of it,
\[
q_{a}^{(1)} \cdot \{\al_{j}\} =
(0,\ldots,0,1,-1,0,\ldots,1,-1,1,-1,\ldots,0,1),
\]
\noi the $j^{th}$ entry of the row vector on the right-hand side is
$q_{a}^{(1)} \cdot \al_{j}$.

It is also elementary to see the following. Suppose a weight
$\gamma_{1}$ has a
scalar product with the simple roots as $\ga_{1} \cdot \{\al_{j}\} =
(0,\ldots,0,1,-1,0,\ldots,0)$. Consider the Weyl reflection
$r$ with respect
to the simple root $\al_{k}$ where $\ga_{1}\cdot \al_{k}=-1$. The action
of $r$ on $\ga_{1}$ will shift the pair $(1,-1)$ in $\ga_{1} \cdot
\{\al_{j}\}$ one step to the right, i.e. $r : \ga_{1} \longrightarrow
\ga_{1}'$ with
\[
\ga_{1}'\cdot \{\al_{j}\} = (0,\ldots,0,0,1,-1,\ldots,0).
\]
\noi For a weight $\ga_{2}$ which has $\ga_{2} \cdot \{\al_{j}\} =
(0,\ldots,0,1,-1,1,\ldots,0)$, consider the Weyl reflection $r'$
with respect to the simple root $\al_{k}$ where $\ga_{2}\cdot\al_{k}=-1$.
The action of $r'$ on $\ga_{2}$ will give $\ga_{2}'$ where,
\[
\ga_{2}'\cdot \{\al_{j}\} = (0,\ldots,0,0,1,0,\ldots,0).
\]

So, using a combination of these Weyl transformations, $q_{a}^{(1)}$ can
be transformed into a fundamental weight, $q_{a}^{(1)} \longrightarrow
\lambda$, where
\[
\lambda \cdot \{\al_{j}\} = (0,\ldots,0,1,0,\ldots,0).
\]
\noi Since there are $(b-1)$ pairs of $(1,-1)$ in $q_{a}^{(1)} \cdot
\{\al_{j}\}$ row vector, then after these combination of Weyl
transformations the entry 1 will appear at the position $n-(b-1)=2a \bmod
h$. Hence the highest topological charge $q_{a}^{(1)}$ lies in the
fundamental representation ${\cal R}_{\lambda_{2a\bmod h}}$.

Recall that the rest of the topological charges are obtained by
cyclically permuting the simple roots and $\al_{0}$, (\ref{q_k}). This
cyclic permutation is the same as the action of  the following Coxeter
element of the Weyl group on $q_{a}^{(1)}$,
\begeq\label{cox}
\omega_{tc} = r_{1}r_{2}\ldots r_{n},
\endeq
\noi where $r_{j}$ is a Weyl reflection with respect to the simple
root $\al_{j}$. Then, the topological charges are related to the highest
charge by,
\begeq\label{cox_on_q}
q_{a}^{(k)} = \omega_{tc}^{k-1} (q_{a}^{(1)}).
\endeq
\noi Note that the ordering of Weyl reflections above is special, other
orderings do not necessarily relate one topological charge to another.
The relation (\ref{cox_on_q}) is straightforward to see using the fact
that,
\begeq\label{cox_roots}
\omega_{tc}(\al_{j}) = \al_{j+1} \qquad \mbox{ for } j=0,1,\ldots,n
\endeq
\noi the simple roots and $\al_{0}$ are labelled modulo $h$.
Further examination of (\ref{q_k}) shows that the set of topological charges
$\{q_{a}^{(k)}\}$ coincides with the topological charges of the species
$2a \bmod h$ single solitons.

The next task is to show that ${\cal R}_{\lambda_{2a\bmod h}}$ is a
component of the Clebsch-Gordan decomposition of ${\cal
R}_{\lambda_{a}} \otimes {\cal R}_{\lambda_{a}}$. This will be shown
using a conjecture attributed to Parthasarathy, Ranga Rao and
Varadarajan \cite{PRV}. The PRV
conjecture may be stated as follows: let $\bar \gamma$ be a unique
dominant weight
of the Weyl orbit of $\gamma = \lambda + \omega\mu$ for any $\omega$
in the Weyl group and $\lambda, \mu$ are highest weights, then ${\cal
R}_{\bar \gamma}$ appears with multiciplity
of  at least one in the decomposition of ${\cal R}_{\lambda}\otimes{\cal
R}_{\mu}$, where ${\cal R}_{\lambda}$ and ${\cal R}_{\mu}$ are finite
dimensional irreducible representations with highest weights $\lambda$ and
$\mu$ respectively.
This conjecture has been proved recently \cite{SK}; it was first
used in the context of affine Toda theories by Braden \cite{HB}.

For convenience of calculation, one can write the fundamental weights of
the Lie algebra $A_n$ as follows,
\begeq\label{lambda_alpha}
\lambda_{a} = \sum_{j=0}^{a} \frac{(h-a)j}{h}\al_{j} + \sum_{j=a+1}^{n}
\frac{a(h-j)}{h}\al_{j}.
\endeq
\noi By the $Z_{2}$ symmetry of the simple roots of $A_n$, one has to
consider only the case $a\leq[\frac{h}{2}]$.
Choose $\omega$ to be the Coxeter element defined in (\ref{cox}).
Then, remembering the action of this Coxeter element on the simple
roots, c.f. (\ref{cox_roots}), it is easy to show that
\begeq
\lambda_{a} + \omega_{tc}^{a}\lambda_{a} = \lambda_{2a}.
\endeq
\noi It is obvious that $\lambda_{2a}$ is a unique dominant weight of
the Weyl orbit. Thus by PRV conjecture ${\cal R}_{\lambda_{2a\bmod h}}
\subset {\cal R}_{\lambda_{a}} \otimes {\cal R}_{\lambda_{a}}$.

This completes the claim that
all the topological charges lie in the same fundamental
representation ${\cal R}_{\lambda_{2a\bmod h}}$ which is an irreducible
component of ${\cal R}_{\lambda_{a}} \otimes {\cal R}_{\lambda_{a}}$,
\begeq
\{q_{a}^{(k)}\} \; \in \; {\cal R}_{\lambda_{2a\bmod h}} \subset {\cal
R}_{\lambda_{a}} \otimes
{\cal R}_{\lambda_{a}}.
\endeq
\noi However, as noted in the
previous calculation, the number of topological charges is $\tilde h =
\frac{h}{\gcd(2a,h)}$ which is generally less than the dimension of
${\cal R}_{\lambda_{2a\bmod h}}$. So, the topological charges of type A
breathers, normally do not fill the fundamental representation ${\cal
R}_{\lambda_{2a\bmod h}}$.
Only particular combinations of the topological charge
of the constituent solitons can make up a breather.
A special case of the
type A breather is when the constituent solitons come from the
fundamental representation ${\cal R}_{\lambda_{a}}$ which is
self-conjugate, this happens for ${\cal R}_{\lambda_{n+1}}$ in the
representation of $A_{2n+1}$. This breather belongs to both type A and B.

For the type B breathers and the exceptional case above, the
fundamental representations of its
constituent solitons are conjugate of each other (or self-conjugate).
Thus, the topological
charge of these breathers will lie in the tensor product ${\cal
R}_{\lambda_{a}} \otimes {\cal R}_{\lambda_{h-a}}$. Using the PRV
conjecture as before, it can be shown that
\begeq
\lambda_{a} + \omega_{tc}^{a}\lambda_{h-a} = 0.
\endeq
\noi Hence, the trivial {\it singlet} representation appears in the
Clebsch-Gordan decomposition of this tensor product.
It is in this singlet
representation that the topological charge lies.

\setcounter{equation}{0}
\section{Sine-Gordon Embedding}
Automorphisms of the Dynkin diagram can be used to reduce an affine Toda
theory to another affine Toda theory with fewer scalar fields \cite{OT3}.
Using this reduction method,
Sasaki noted that in the \an affine Toda theories with a real coupling
parameter, there are ways to reduce some members of the \an family to the
$a_{1}^{(1)}$ theory,
i.e. the sinh-Gordon theory \cite{RS}. The same procedure can be applied in
the case of complex Toda theories.
Define the solution to the equation of motion (\ref{eom}) as,
\begeq\label{ansatz}
\phi = \mu \psi,
\endeq
\noi where $\mu$ is some vector to be determined. Then (\ref{eom}) becomes,
\begeq\label{sg_eom1}
\mu \partial^{2}(\be \psi) = i m^{2} \sum_{j=1}^{n} \al_{j} \left(e^{i\be
\al_{j}\cdot\mu\psi} - e^{i\be\al_{0}\cdot\mu\psi}\right).
\endeq
The aim is to reduce (\ref{sg_eom1}) above into the sine-Gordon equation
of motion by choosing a suitable $\mu$,
\begeq\label{sg_eom2}
\mu \partial^{2}(\be \psi) = i m^{2} \mu \left(
e^{i\be\psi}-e^{-i\be\psi} \right) = -2m^{2} \mu \sin(\be\psi).
\endeq

There are two kinds of reductions. A {\it direct reduction} results when
several nodes of the affine Dynkin diagram which do not have a direct link
are identified. When linked nodes are transposed, this results in a
{\it non-direct reduction}.

One can reduce the $a_{2n+1}^{(1)}$ theories to $a_{1}^{(1)}$ theory
using a direct reduction by
choosing $\mu$ as follows \cite{RS},
\begeq
\mu_1 = \al_{1} + \al_{3} + \ldots + \al_{2n-1} + \al_{2n+1}.
\endeq
\noi The vector $\mu_1$ is an invariant vector under the $Z_{n+1}$
symmetry which identifies $\al_{j} \rightarrow \al_{j+2}$. Projecting the
simple roots of $a_{2n+1}^{(1)}$ to $\mu_{1}$
subspace gives the simple roots of $a_{1}^{(1)}$ with multiplicity $(n+1)$,
\[
\al_{j}\cdot\mu_{1} = 2 \mbox{ or } -2,
\]
\noi for $j$ odd or even respectively.
A {\it non-direct reduction} is a two step
reduction, first $a_{4n-1}^{(1)}$ can be reduced to the three dimensional
subspace of $a_{3}^{(1)}$ then $a_{3}^{(1)}$ can be reduced to
$a_{1}^{(1)}$ \cite{RS}. There are two choices of $\mu$
for this non-direct reduction,
\begin{eqnarray}
\mu_{2} = \al_{1} + \al_{2} + \al_{5} + \al_{6} + \ldots + \al_{4n-3} +
\al_{4n-2} ,\\
\mu_{3} = \al_{2} + \al_{3} + \al_{6} + \al_{7} + \ldots + \al_{4n-2} +
\al_{4n-1},
\end{eqnarray}
\noi in the above, $\mu_3$ is obtained from $\mu_{2}$ by cyclically
permuting the simple roots of $a_{4n-1}^{(1)}$ once. Together with the
vector $\mu_{1}$, these three vectors are invariant under the $Z_{n}$
symmetry which identifies $\al_{j} \rightarrow \al_{j+4}$. The simple
roots of $a_{4n-1}^{(1)}$ can be projected to $\mu_{2}$ or $\mu_{3}$
giving the simple roots of $a_{1}^{(1)}$ with multiplicity $2n$.

In terms of the single soliton $\tau$-functions (\ref{1_sol}), direct
reduction forces some $\tau$-functions to be equal leaving only two
different $\tau$-functions,
\begeq\label{direct_tau_cond}
\tau_{0}^{(a)}=\tau_{2}^{(a)}=\ldots=\tau_{2n}^{(a)} \qquad \mbox{ and }
\qquad \tau_{1}^{(a)}=\tau_{3}^{(a)}=\ldots=\tau_{2n-1}^{(a)},
\endeq
\noi with,
\[
\tau_{j}^{(a)} = 1 + \omega_{a}^{j}e^{(\Omega_{a}+\rho)}.
\]
\noi Since $\omega_{a}^{j} = \exp(\frac{2i\pi a}{h}j)$, it is  clear
that for the $a_{2n+1}^{(1)}$ theories, only solitons of species $a=(n+1)$
are the {\it true} sine-Gordon solitons embedded in the theory.
For the non-direct reductions, one has to have the following condition for
the $\tau$-functions. Using $\mu_{2}$ yields,
\begin{eqnarray*}
\tau_{0}^{(a)}=\tau_{3}^{(a)}=\tau_{4}^{(a)}=
\ldots =\tau_{4n-4}^{(a)}=\tau_{4n-1}^{(a)} ,\\
\tau_{1}^{(a)}=\tau_{2}^{(a)}=\tau_{5}^{(a)}= \ldots
=\tau_{4n-3}^{(a)}=\tau_{4n-2}^{(a)},
\end{eqnarray*}
\noi and for the choice of $\mu_{3}$,
\begin{eqnarray*}
\tau_{0}^{(a)}=\tau_{1}^{(a)}=\tau_{4}^{(a)}=
\ldots =\tau_{4n-4}^{(a)}=\tau_{4n-3}^{(a)} , \\
\tau_{2}^{(a)}=\tau_{3}^{(a)}=\tau_{6}^{(a)}= \ldots
=\tau_{4n-2}^{(a)}=\tau_{4n-1}^{(a)}.
\end{eqnarray*}
\noi These conditions on the $\tau$-functions of $a_{4n-1}^{(1)}$ will
never be satisfied. This is because for $h=4n$, the factor $\omega_{a}^{j}$
cannot be equal to $\omega_{a}^{j+1}$ since $j$ and $(j+1)$ are coprime.

Thus, the solitons associated with middle spot of the $A_{2n+1}$ Dynkin
diagram are the only sine-Gordon solitons embedded in the
$a_{2n+1}^{(1)}$ affine Toda theories. Hence, these solitons can bind
together resulting in sine-Gordon breathers, i.e. type A breathers with
zero topological charge.
Note also that type B breathers by the above definitions are not formed
{}from any sine-Gordon embedded solitons.

\setcounter{equation}{0}
\section{Examples}
In this section the case of $a_{3}^{(1)}$ and $a_{4}^{(1)}$ breathers
will be given as examples.

The number above each spot on the Dynkin diagram are the number of
topological charges of the type A breathers constructed from solitons
associated with each spot (see figure 2).
The topological charges of the type A breather, $q_{a}^{(k)}$, are listed
below. The subscript denotes the species of the constituent solitons and
the
superscript labels the topological charges, $q_{a}^{(1)}$ is the {\it
highest} topological charge.
\begin{eqnarray*}
q_{1}^{(1)} &=& \frac{1}{2}\al_{1} + \frac{1}{2}\al_{3} ,\\
q_{1}^{(2)} &=& -\frac{1}{2}\al_{1} - \frac{1}{2}\al_{3},
\end{eqnarray*}
\noi and,
\[
q_{2}^{(1)} = 0.
\]
\noi The topological charges $\{q_{3}\}$ is the same as $\{q_{1}\}$ and not
listed above.
All type B breathers have zero topological charge.

The topological charges $\{q_{1}\}$, and hence also $\{q_{3}\}$,
lie in the second fundamental representation, ${\cal R}_{\lambda_{2}}
\subset {\cal R}_{\lambda_{1}} \otimes {\cal
R}_{\lambda_{1}}$ or ${\cal R}_{\lambda_{3}} \otimes {\cal
R}_{\lambda_{3}}$. The dimension of ${\cal R}_{\lambda_{2}}$ is 6, and
there are only 2 topological charges for $\phi_{11}$ or $\phi_{33}$
breathers. Thus, these topological charges do not fill up ${\cal
R}_{\lambda_{2}}$.
For $q_{2}$, as explained in the previous section,
this is an embedded sine-Gordon breather, hence $q_{2}=0$.

Since topological charges are conserved quantities, it follows that
for both type A breathers and type B breathers $q_{a}$ is equal to
the sum  of the topological charges of
the constituent solitons. Thus only a special combination of constituent
solitons can make up a breather.

The Dynkin diagram for $A_{4}$ can be found in figure 3. The
topological charges of type A breathers in $a_{4}^{(1)}$ are listed as
follows. Breathers from species $a=1$ solitons:
\begin{eqnarray*}
q_{1}^{(1)} &=& \frac{3}{5}\al_{1} + \frac{1}{5}\al_{2} + \frac{4}{5}\al_{3}
+ \frac{2}{5}\al_{4} ,\\
q_{1}^{(2)} &=& -\frac{2}{5}\al_{1} + \frac{1}{5}\al_{2} - \frac{1}{5}\al_{3}
+ \frac{2}{5}\al_{4} ,\\
q_{1}^{(3)} &=& -\frac{2}{5}\al_{1} - \frac{4}{5}\al_{2} - \frac{1}{5}\al_{3}
- \frac{3}{5}\al_{4}, \\
q_{1}^{(4)} &=& \frac{3}{5}\al_{1} + \frac{1}{5}\al_{2} - \frac{1}{5}\al_{3}
+ \frac{2}{5}\al_{4}, \\
q_{1}^{(5)} &=& -\frac{2}{5}\al_{1} + \frac{1}{5}\al_{2} - \frac{1}{5}\al_{3}
- \frac{3}{5}\al_{4}.
\end{eqnarray*}
\noi Breathers from species $a=2$ solitons:
\begin{eqnarray*}
q_{2}^{(1)} &=& \frac{1}{5}\al_{1} + \frac{2}{5}\al_{2} + \frac{3}{5}\al_{3}
+ \frac{4}{5}\al_{4} ,\\
q_{2}^{(2)} &=& -\frac{4}{5}\al_{1} - \frac{3}{5}\al_{2} - \frac{2}{5}\al_{3}
- \frac{1}{5}\al_{4} ,\\
q_{2}^{(3)} &=& \frac{1}{5}\al_{1} - \frac{3}{5}\al_{2} - \frac{2}{5}\al_{3}
- \frac{1}{5}\al_{4} ,\\
q_{2}^{(4)} &=& \frac{1}{5}\al_{1} + \frac{2}{5}\al_{2} - \frac{2}{5}\al_{3}
- \frac{1}{5}\al_{4} ,\\
q_{2}^{(5)} &=& \frac{1}{5}\al_{1} + \frac{2}{5}\al_{2} + \frac{3}{5}\al_{3}
- \frac{1}{5}\al_{4}.
\end{eqnarray*}
\noi Breathers from species $a=3$ solitons:
\begin{eqnarray*}
q_{3}^{(1)} &=& \frac{4}{5}\al_{1} + \frac{3}{5}\al_{2} + \frac{2}{5}\al_{3}
+ \frac{1}{5}\al_{4} ,\\
q_{3}^{(2)} &=& -\frac{1}{5}\al_{1} + \frac{3}{5}\al_{2} + \frac{2}{5}\al_{3}
+ \frac{1}{5}\al_{4} ,\\
q_{3}^{(3)} &=& -\frac{1}{5}\al_{1} - \frac{2}{5}\al_{2} + \frac{2}{5}\al_{3}
+ \frac{1}{5}\al_{4} ,\\
q_{3}^{(4)} &=& -\frac{1}{5}\al_{1} - \frac{2}{5}\al_{2} - \frac{3}{5}\al_{3}
+ \frac{1}{5}\al_{4} ,\\
q_{3}^{(5)} &=& -\frac{1}{5}\al_{1} - \frac{2}{5}\al_{2} - \frac{3}{5}\al_{3}
- \frac{4}{5}\al_{4}.
\end{eqnarray*}
\noi Breathers from species $a=4$ solitons:
\begin{eqnarray*}
q_{4}^{(1)} &=& \frac{2}{5}\al_{1} + \frac{4}{5}\al_{2} + \frac{1}{5}\al_{3}
+ \frac{3}{5}\al_{4} ,\\
q_{4}^{(2)} &=& -\frac{3}{5}\al_{1} - \frac{1}{5}\al_{2} + \frac{1}{5}\al_{3}
- \frac{2}{5}\al_{4} ,\\
q_{4}^{(3)} &=& \frac{2}{5}\al_{1} - \frac{1}{5}\al_{2} + \frac{1}{5}\al_{3}
+ \frac{3}{5}\al_{4} ,\\
q_{4}^{(4)} &=& -\frac{3}{5}\al_{1} - \frac{1}{5}\al_{2} - \frac{4}{5}\al_{3}
- \frac{2}{5}\al_{4} ,\\
q_{4}^{(5)} &=& \frac{2}{5}\al_{1} - \frac{1}{5}\al_{2} + \frac{1}{5}\al_{3}
- \frac{2}{5}\al_{4}.
\end{eqnarray*}
\noi These topological charges lie in the following fundamental
representation,
\begin{eqnarray*}
\{q_{1}^{(k)}\} \; &\in & \; {\cal R}_{\lambda_{2}} \subset {\cal
R}_{\lambda_{1}} \otimes {\cal R}_{\lambda_{1}} ,\\
\{q_{2}^{(k)}\} \; &\in & \; {\cal R}_{\lambda_{4}} \subset {\cal
R}_{\lambda_{2}} \otimes {\cal R}_{\lambda_{2}} ,\\
\{q_{3}^{(k)}\} \; &\in & \; {\cal R}_{\lambda_{1}} \subset {\cal
R}_{\lambda_{3}} \otimes {\cal R}_{\lambda_{3}} ,\\
\{q_{4}^{(k)}\} \; &\in & \; {\cal R}_{\lambda_{3}} \subset {\cal
R}_{\lambda_{4}} \otimes {\cal R}_{\lambda_{4}}.
\end{eqnarray*}
There is no sine-Gordon embedding in this case.
And the topological charges of all type B breathers are zero.

\section{Conclusions}
Following the example of the sine-Gordon theory, classical oscillating
soliton
solutions of the \an affine Toda theories have been constructed as
bound states of soliton pairs. These breathers are classified by the
species of the constituent solitons. These can either be two solitons of
the same species (type
A breathers) or solitons of antispecies (type B breathers).

The topological charges of these breather solutions have been calculated.
Type A breathers carry topological charges which lie in the
fundamental representation ${\cal R}_{\lambda_{2a\bmod h}}$ $\subset$ ${\cal
R}_{\lambda_{a}} \otimes {\cal R}_{\lambda_{a}}$, where $a$ is the
species of the constituent solitons. To be precise, these topological charges
coincide with the
topological charges of the single soliton of species $2a \bmod h$.
Therefore the fundamental representation ${\cal R}_{\lambda_{2a\bmod
h}}$ is normally not filled up \cite{WAM}. It is a mystery that only certain
combinations of the topological
charges of the constituent solitons are allowed to bind together
to yield a breather.
An understanding of these phenomena is far from complete.
It is conjectured that in the quantum theory there are more states
than classical
solutions \cite{TH2}. In other words, the topological
charges in the quantum theory have been
conjectured to fill up the associated fundamental representation.
As part of the spectrum of the quantum theory corresponds to classical
soliton and breather solutions, one might have thought that the
classical breather solutions give at least some of the missing
topological charges. This appears  not to be the case, at least for
breathers with two constituent solitons.

Exceptional cases of type A breathers are those
constructed from solitons of species $(n+1)$ in the $a_{2n+1}^{(1)}$
theories. These are  embedded sine-Gordon breathers
which belong to type A and type B  since
both carry zero topological charge. They differ from the type B breathers
as type B are sine-Gordon {\it like} breathers.
It has been shown that in  both of these cases, the topological
charge lies in the {\it singlet} component of the Clebsch-Gordan
decomposition of the tensor product of a
fundamental representation with its conjugate representation.

Of no less important interest are breathers in
other affine
Toda theories. It will be interesting to know how many breathers can be
constructed from their solitons.

\section*{Acknowledgements}
We would like to thank E. Corrigan for helpful discussions and
encouragement in tackling the problems. We
would also like to acknowledge P. E. Dorey and R. Hall for their comments
and suggestions.
UH would like to thank the Commission of European Communities for a
grant given  under the Human Capital and Mobility programme (grant no.
ERB4001GT920871).
AAI would like to thank the Ministry of Education and Culture
(Indonesia) for a studentship given under the second Higher Education
Development Project and
the UK Committee of Vice-Chancellors and Principals for an ORS award.
WAM would like to thank the UK Science and Engineering Research Council
for a studentship.

\clearpage
%%%%%%%%%%%%%%%%%%%%%%%%%%%%%%%% REFERENCES %%%%%%%%%%%%%%%%%%%%%%%%%%%%%%%%%%%
%

\clearpage

\listoffigures

\clearpage
\setcounter{figure}{-1}
\begin{figure}[h]
 \begin{minipage}{0cm} \end{minipage}
 \hfill
 \begin{minipage}{10cm}
 \refstepcounter{figure}\label{pic}
 \epsfxsize=10cm
 \epsffile{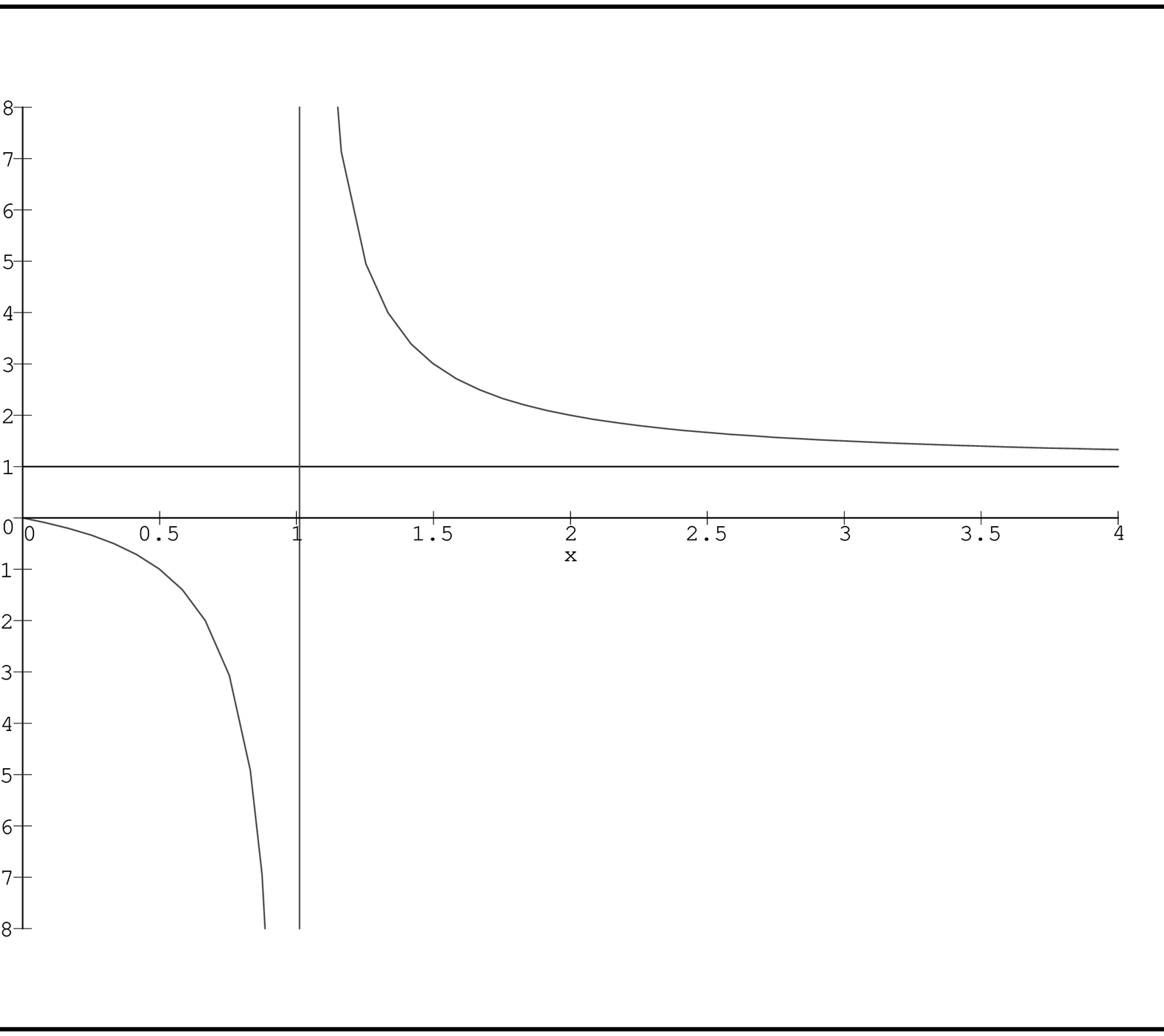}
 \end{minipage}
 \hfill
 \begin{minipage}{0cm} \end{minipage}
\caption{The graph shows
the behaviour of the interaction coefficient for the type A breather
solution. The interaction $A'$ is given by $A'= A \cos(\theta/2)$, the
velocity as $x= v^2 / v^2_c$. As mentioned in section 3.1 there is
an upper bound for the velocity. This has not been marked in the
graph.}
 \end{figure}
\renewcommand{\thefootnote}{}
\footnote{{\bf On the Breathers of \an Affine Toda Field Theory,}
by {\tt Uli Harder, Alexander A. Iskandar, William A. McGhee}}

\clearpage
\begin{figure}
\vspace{8cm}
{\centering
\begin{picture}(180,30)(11,50)
\put ( 208,50){\line( 1, 0){24}}
\put ( 248,50){\line( 1, 0){24}}
\put ( 200,50){\circle{10}}
\put ( 240,50){\circle{10}}
\put ( 280,50){\circle{10}}
\put (200,65){\makebox(0,0){$2$}}
\put (240,65){\makebox(0,0){$1$}}
\put (280,65){\makebox(0,0){$2$}}
\end{picture}}
\caption{Dynkin diagram for $A_3$.}
\end{figure}

\footnote{{\bf On the Breathers of \an Affine Toda Field Theory,}
by {\tt Uli Harder, Alexander A. Iskandar, William A. McGhee}}

\clearpage
\begin{figure}
\vspace{8cm}
{\centering
\begin{picture}(180,30)(-20,50)
\put ( 158,50){\line( 1, 0){24}}
\put ( 198,50){\line( 1, 0){24}}
\put ( 238,50){\line( 1, 0){24}}
\put ( 150,50){\circle{10}}
\put ( 190,50){\circle{10}}
\put ( 230,50){\circle{10}}
\put ( 270,50){\circle{10}}
\put (150,65){\makebox(0,0){$5$}}
\put (190,65){\makebox(0,0){$5$}}
\put (230,65){\makebox(0,0){$5$}}
\put (270,65){\makebox(0,0){$5$}}
\end{picture}}
\caption{Dynkin diagram for $A_4$.}
\end{figure}
\footnote{{\bf On the Breathers of \an Affine Toda Field Theory,}
by {\tt Uli Harder, Alexander A. Iskandar, William A. McGhee}}


\vspace{0.3in}
\baselineskip 18pt
\begin{thebibliography}{199}


\bibitem{AFZ+MOP}
A.E. Arinshtein, V.A. Fateev and A.B. Zamolodchikov, {\it Phys. Lett.}
{\bf B87} (1979), 389. \\
A.V. Mikhailov, M.A. Olshanetsky and A.M. Perelomov, {\it Comm. Math.
Phys.} {\bf 79} (1981), 473.

\bibitem{OT3}
D.I. Olive and N. Turok, {\it Nucl. Phys.} {\bf B215} (1983), 470.

\bibitem{OT1+2}
D.I. Olive and N. Turok, {\it Nucl. Phys.} {\bf B220} (1983), 491. \\
D.I. Olive and N. Turok, {\it Nucl. Phys.} {\bf B257} (1985), 277.

\bibitem{HM+EY}
T. Eguchi and S.-K. Yang, {\it Phys. Lett} {\bf B224} (1989), 373. \\
T. Hollowood and P. Mansfield, {\it Phys. Lett.} {\bf B226} (1989), 73.

\bibitem{BCDS1}
H.W. Braden, E. Corrigan, P.E. Dorey and R. Sasaki, {\it Nucl. Phys.} {\bf
B338} (1990), 689.

\bibitem{CM}
P. Christe and G. Mussardo, {\it Nucl. Phys.} {\bf B330} (1990), 465. \\
P. Christe and G. Mussardo, {\it Int. J. Mod. Phys.} {\bf A5} (1990), 4581.

\bibitem{BCDS2}
H.W. Braden, E. Corrigan, P.E. Dorey and R. Sasaki, {\it Nucl. Phys.} {\bf
B356} (1991), 469.

\bibitem{PED}
P.E. Dorey, {\it Nucl. Phys.} {\bf B358} (1991), 654. \\
P.E. Dorey, {\it Nucl. Phys.} {\bf B374} (1992), 741.

\bibitem{MF+FLO}
M.D. Freeman, {\it Phys. Lett.} {\bf B261} (1991), 57. \\
A. Fring, H.C. Liao and D.I. Olive, {\it Phys. Lett.} {\bf B266} (1991), 82.

\bibitem{BS+SZ}
H.W. Braden and R. Sasaki, {\it Phys. Lett.} {\bf B255} (1991), 343. \\
H.W. Braden and R. Sasaki, {\it Nucl. Phys.} {\bf B379} (1992), 377. \\
R. Sasaki and F.P. Zen, {\it Int. J. Mod. Phys.} {\bf A8} (1993), 115.

\bibitem{NSL}
G.W. Delius, M.T. Grisaru and D. Zanon, {\it Nucl. Phys.} {\bf B382}
(1992), 365. \\
E. Corrigan, P.E. Dorey and R. Sasaki, {\it Nucl. Phys.} {\bf B408}
(1993), 579. \\
P.E. Dorey, {\it Phys. Lett.} {\bf B312} (1993), 291.

\bibitem{TH1}
T. Hollowood, {\it Nucl. Phys.} {\bf B384} (1992), 523.

\bibitem{OTU1}
D.I. Olive, N. Turok and J.W.R. Underwood, {\it Nucl. Phys.} {\bf B401}
(1993), 663.

\bibitem{OTU2}
D.I. Olive, N. Turok and J.W.R. Underwood, {\it Nucl. Phys.} {\bf B409}
(1993), 509.

\bibitem{FJKO}
A. Fring, P.R. Johnson, M.A.C. Kneipp and D.I. Olive, {\it  Vertex
Operators and Soliton Time Delays in Affine Toda Field
Theory}, SWAT-93-94-30, hep-th/9405034.

\bibitem{TH2}
T. Hollowood, {\it Int. Jour. Mod. Phys.} {\bf A8} (1993), 947.

\bibitem{LS}
A.N. Leznov and M.V. Saveliev, {\it Lett. Math. Phys.} {\bf 3} (1979), 485.

\bibitem{PM}
P. Mansfield, {\it Comm. Math. Phys.} {\bf 98} (1985), 525.

\bibitem{MM}
N.J. MacKay and W.A. McGhee, {\it Int. J. Mod. Phys.} {\bf A8} (1993), 2791.

\bibitem{Brasil}
C.P. Constantinidis, L.A. Ferreira, J.F. Gomes and A.H. Zimerman, {\it Phys.
Lett.} {\bf B298} (1993), 88. \\
H. Aratyn, C.P. Constantinidis, L.A. Ferreira, J.F. Gomes and A.H. Zimerman,
{\it Nucl. Phys.} {\bf B406} (1993), 727.

\bibitem{Raja}
R. Rajaraman, {\it Instanton and Solitons}, North Holland, (1982).

\bibitem{RH}
R. Hirota, {\it Direct methods in soliton theory}, in {\it Solitons},
eds. R.K. Bullogh and P.S. Caudrey (1980), p. 157.

\bibitem{RAH}
R.A. Hall, Ph.D. thesis, Durham University, 1994; unpublished.

\bibitem{WAM}
W.A. McGhee, {\it Int. J. Mod. Phys.} {\bf A9} (1994), 2645.

\bibitem{PRV}
K.R. Parthasarathy, R. Ranga Rao and V.S. Varadarajan, {\it Ann. Math.} {\bf
85} (1967), 383.

\bibitem{SK}
S. Kumar, {\it Invent. Math.} {\bf 93} (1988), 117.

\bibitem{HB}
H.W. Braden, {\it J. Phys.} {\bf A25} (1992), L15.

\bibitem{RS}
R. Sasaki, {\it Nucl. Phys.} {\bf B383} (1992), 291.

\end{thebibliography}
\end{document}